\documentclass[aps,prl,twocolumn]{revtex4}
\usepackage{epsfig}
\usepackage{color}
\usepackage{epstopdf}

\begin{document}               

\title{Finite-size scaling for discontinuous nonequilibrium phase
transitions}

\author{Marcelo M. de Oliveira$^{1}$, M. G. E. da Luz$^2$ and 
Carlos E. Fiore$^3$}
\address{
$^1$Departamento de F\'{\i}sica e Matem\'atica,
CAP, Universidade Federal de S\~ao Jo\~ao del Rei,
36420-000 Ouro Branco-MG, Brazil,
\\
$^2$ 
Departamento de F\'isica, Universidade Federal do Paran\'a, 
81531-980 Curitiba-PR, Brazil \\
$^3$ Instituto de F\'isica, Universidade de S\~ao Paulo, 
05315-970 S\~ao Paulo-SP, Brazil
}

\date{\today}

\begin{abstract}
A finite size scaling theory, originally developed only for transitions
to absorbing states [Phys. Rev. E {\bf 92}, 062126 (2015)], is extended
to distinct sorts of discontinuous nonequilibrium phase transitions.
Expressions for quantities such as, response functions, reduced cumulants
and equal area probability distributions, are derived from phenomenological
arguments.
Irrespective of system details, all these quantities scale with the
volume, establishing the dependence on size.
The approach generality is illustrated through the analysis of different
models.
The present results are a relevant step in trying to unify the
scaling behavior description of nonequilibrium transition processes.
\end{abstract}


\maketitle


Methodologically speaking, nonequilibrium bears a relation to
equilibrium phase transitions somewhat similar to that between nonlinear
and linear dynamical systems.
Indeed, well established theoretical frameworks, universal methods of
analysis and generic efficient calculation protocols are far better
developed for the latter than for the former.
Yet, nonequilibrium are conceivably more prevalent than equilibrium
transitions, ubiquitous in such diverse phenomena 
\cite{marro,odor07,henkel,hinrichsen,odor04} as, interface growth,
epidemics, chemical reactions, population dynamics, flow in biological
systems \cite{niven-2010}, spatio-temporal chaos in liquid crystals
\cite{take07}, behavior of driven suspensions \cite{pine}, and
superconducting vortices \cite{okuma} (see \cite{take14} for a survey).

Most of the studies on nonequilibrium has been directed to
universality and scaling in the continuous context, with much less
attention being paid to discontinuous nonequilibrium phase transitions
(DNPT) \cite{fiore14,fss15}.  
Despite this, DNPT are quite commonly observed in problems like
heterogeneous catalysis 
\cite{ehsasi,zgb},  ecological processes \cite{scp,deser}, granular 
systems \cite{soto}, replicator dynamics \cite{fontanari}, cooperative 
coinfection \cite{coinfection}, language formation \cite{naming} 
and social influence \cite{castellano}.

There are few general results for nonequilibrium phase transitions
\cite{odor07,henkel,kondepudi-1998}, one example being the celebrated
Jarzynski equality \cite{jarzynski-1997} for free energy differences
$\Delta F$. 
Another, for DNPT, is that to emerge they seem to require an effective
mechanism suppressing the formation of minority islands induced
by fluctuations \cite{lubeck,grassberger}. 
Also, DNPT allow a finite-size scaling (FSS).
But this fact has been shown only for transitions to absorbing states
\cite{fss15}, based on the idea of quasi-stationary (QS) ensembles
\cite{dickman-vidigal02,qssim}.
Calculations for several cases in $d$ dimensions reveal
that distinct QS quantities scale with the system volume $V = L^d$.
Further, the shift in the coexistence point value goes with $1/V$.
Such findings are in great similarity with equilibrium phase transitions
\cite{fisher,binder, binder2,binder3,lee,borgs,landau,fioreprl},
whose coexistence points correspond to equal $F$'s in the distinct 
phases and for the $F$'s second derivatives displaying a
$\delta$-function-like shape in the thermodynamic limit.

Here we extend the results in \cite{fss15} for systems admitting
nonequilibrium steady states (without exhibiting absorbing states).
Relying on solid phenomenological arguments, we deduce a general FSS
behavior for DNPT.
We determine that different quantities, like response functions,
reduced ratio cumulants and probability distributions, go with the
inverse of the system volume.
We illustrate our method addressing diverse nonequilibrium models, as 
absorption-desorption on catalytic surfaces, systems with up-down $Z_2$ 
symmetry, and competitive interactions in bipartite sublattices, all them
displaying distinct features and symmetries.
Our approach unveils an universal scaling behavior for DNPT in close
analogy with the equilibrium case.
Hence, it constitutes an additional example of the unexpected resemblance 
between procedures treating certain aspects of equilibrium and nonequilibrium 
thermodynamics \cite{fss15,sivak-2012}.


Our starting point is that in some relevant aspects equilibrium and 
nonequilibrium phase transitions are akin when the latter present
stationary states \cite{similarities}.
For instance, as very clearly put in \cite{landsberg-1980}, for a
sufficiently large variation of a control parameter $\lambda$,
DNPT can be viewed as the passage through distinct steady states,
each typified by the value of an order parameter $m$.
Moreover --- at least regarding $m$ --- these states must be stable
against \cite{kondepudi-1998} external disturbances, fluctuations,
and eventual internal currents (typically arising when the detailed
balance is not observed).

The above is the case when the net probability current (yielding
the order parameter time evolution) in each steady state is null, i.e.,
if the microscopic dynamics always obeys the global balance, namely
\cite{gb-ss},
\begin{equation}
\sum_{y} p(x) \, w(x \rightarrow y) = \sum_{y} p(y) \, w(y \rightarrow x),
\label{eq:global-balance}
\end{equation}
for $x, y$ micro-configurations associated to $m$, $w(x \rightarrow y)$
their transition rate, and $p(z)$ the probability of $z$.
Therefore, in such situation we can ascribe a stable probability density 
to the nonequilibrium steady state \cite{stable-pd}.
Also, its stationarity allied to the global balance support an 
extended version of the central limit theorem for the distribution of
$m$, given by Gaussians \cite{gaussian}, even in the nonequilibrium
case \cite{sivak-2012}.


Thus, in the vicinity of the coexistence at $\lambda_0$ (i.e., for
$\Delta \lambda = \lambda - \lambda_0$ small) and for a finite but 
large enough volume $V$, the phase $C$ probability distribution 
$P_{V}^{(C)}(m)$ should read (up to a normalization constant, see below)
\begin{equation} 
P_{V}^{(C)}(m) =  \sqrt{\frac{g(V)}{2 \pi}}
\exp\Big[g(V) 
\Big(\Delta \lambda \, m - \frac{(m-m_C)^2}{2 \chi_C} \Big)\Big].
\label{eq1} 
\end{equation}
Here $\chi_C$ is the variance and $g(V)$ an increasing function of $V$.
For $V \rightarrow \infty$, Eq. (\ref{eq1}) leads to a $\delta$-function 
centered at $m=m_C$ when $\Delta \lambda = 0$.
In the following, we will consider the extensive case, so that
$g(V) = V$.

At the coexistence of two phases, $A$ ($\lambda \leq \lambda_0$) and $B$
($\lambda \geq \lambda_0$), we have the full probability distribution
\begin{equation}
  P_{V}(m) = \frac{P_{V}^{(A)}(m) + P_{V}^{(B)}(m)}
  {[F_A(\Delta \lambda;V) + F_B(\Delta \lambda;V)]},
  \label{eq2}
\end{equation}
where the denominator in Eq. (\ref{eq2}) gives the correct normalization,
with 
\begin{equation}
F_C(\Delta \lambda;V) = \sqrt{\chi_C} \, \exp\Big[V \Big( \Delta \lambda \,
{m_C} +  \frac{\chi_C}{2}  \Delta \lambda^2
\Big)\Big].
\label{eq3}
\end{equation}

It is simple to analytically calculate any moment of the order parameter
$m$, defined as
\begin{equation}
\langle m^n \rangle_V = \int_{-\infty}^{+\infty} dm \,
m^n \, P_V(m),
\end{equation}
by means of Eqs. (\ref{eq1})--(\ref{eq3}) and the formula
\cite{weissten-2003}
\begin{eqnarray}
\int_{-\infty}^{+\infty} dx \, x^n \exp[-a x^2 + b x] &=&
\sqrt{\frac{\pi}{a}} \exp\Big[\frac{b^2}{4 a}\Big]
\, \sum_{k = 0}^{k = n-1}
\nonumber \\
& & \times \frac{n!}{k! (n - 2 k)!}
\frac{(2 b)^{n - 2 k}}{(4 a)^{n-k}}. \nonumber \\
\end{eqnarray}
In particular, we readily obtain closed analytic expressions for
(our purposes very useful) second and fourth order reduced cumulants,
namely, $U_2 = \langle m^2 \rangle_V/\langle m \rangle^{2}_V$ and
$U_4 = 1 - \langle m^4 \rangle_V/(3 \, \langle m^2\rangle_V)$.

The pseudo-transition point $\lambda = \lambda_V$ can be estimated in
different ways, such as for both phases presenting equal weights (areas)
or through the maximum of the global variance
$V (\langle m^2 \rangle_V - {\langle m \rangle}_V^2)$.
From the present formalism, we have shown in \cite{fss15} that for
$V$ large and in first order in $(\lambda_V - \lambda_0)$, one gets 
in both cases
\begin{equation}
  \lambda_V  \approx  \lambda_0 - \frac{1}{2 V} \,
  \frac{\ln[\chi_B/\chi_A]}{(m_B - m_A)}.
\label{eq4}
\end{equation}
On the other hand, seeking for a $\lambda_V$ given the maximum of $U_2$,
one finds \cite{fss15}
\begin{equation}
\label{eq:basicsolcum}
\lambda_V \approx \lambda_0 - \frac{1}{2 V} \,
\frac{\ln[\chi_B/\chi_A] + 2 \ln[m_B/m_A] }{(m_B - m_A)}.
\label{eq5}
\end{equation}
As we will see, Eqs. (\ref{eq4}) and (\ref{eq5}) give
very similar results for large $V$'s
(comparing with Eqs. (\ref{eq4}) and (\ref{eq5}), we should
mention a small misprint in the denominator of the corresponding Eqs.
in \cite{fss15}).
Likewise DNPT to absorbing states \cite{fss15}, the difference 
between $\lambda_V$ and $\lambda_0$ consistently scales with $1/V$ 
(this is also true by calculating $\lambda_V - \lambda_0$ 
from the minimum of $U_4$, however an expression not shown here).
In the following we demonstrate the usefulness of our analysis by 
discussing distinct models.



{\em The ZGB model with desorption of CO}:
As a first example, we address the Ziff-Gulari-Barshard (ZGB) model
\cite{zgb} with CO desorption \cite{brosilow,tome}.
The original ZGB is a longstanding model reproducing relevant features
of carbon monoxide oxidation on a catalytic surface
(represented by a square lattice).
Each site can be empty ($*$) or occupied by either an oxygen, O, or a carbon
monoxide, CO. 
The interaction rules are:
\begin{eqnarray}
\mbox{CO}_{{\mbox{\scriptsize gas}}} + * \to \mbox{CO}_{{\mbox{\scriptsize ads}}}
\nonumber \\
{\mbox{O}_{2}}_{{\mbox{\scriptsize gas}}} + 2 \, * \to
2 \mbox{O}_{{\mbox{\scriptsize ads}}} \nonumber \\
\mbox{CO}_{{\mbox{\scriptsize ads}}} + \mbox{O}_{{\mbox{\scriptsize ads}}} \to
\mbox{CO}_{2} + 2 *. \nonumber
\end{eqnarray}
\noindent 
Molecules of CO (O$_2$) reach the surface with probability $Y$ ($1-Y$).
Whenever a CO molecule hits a vacant site, the site becomes occupied. 
At the surface, if the O$_2$ molecule finds two vacant nearest-neighbor
sites, each O occupies one of them.
If a CO$_{2}$ molecule is formed (when CO and O are neighbors), it is
immediately desorbed.

In the limit of large (low) $Y$'s, the ZGB exhibits phase transitions
between active steady and absorbing states, with the surface being
saturated, poisoned, by molecules of CO (O).
In the former (latter) situation we have strong discontinuous (continuous,
in the DP universality class) transitions.
This first-order absorbing phase transition has been analyzed in Ref.
\cite{fss15}, and in fact has attracted great theoretical interest
\cite{zgbqs}.

Experimentally, however, one observes a transition from low to high
densities of CO, $\rho_{\mbox{\scriptsize CO}}$, in the lattice, but without
complete poisoning \cite{fisher2,evans,zhdanov}.
This is a consequence of CO desorption, governed by the substrate
temperature.
Above a certain critical temperature, $\rho_{\mbox{\scriptsize CO}}$ varies
smooth and there is no phase transition.
The inclusion of a CO desorption rate $k$ in the initial ZGB model
turns out to be very effective in describing such effect.
Actually, phases of high and low $\rho_{\mbox{\scriptsize CO}}$ arises only
for $k$ less than some critical value
$k_c\approx 0.039$ \cite{brosilow,tome}.
The desorption (an extra step added to the previous set of rules)
is simply implemented by randomly choosing a site,
and if in it there is a CO molecule, then with probability $k$ the
CO is desorbed.

\begin{figure}[!t]
\includegraphics[scale=0.31]{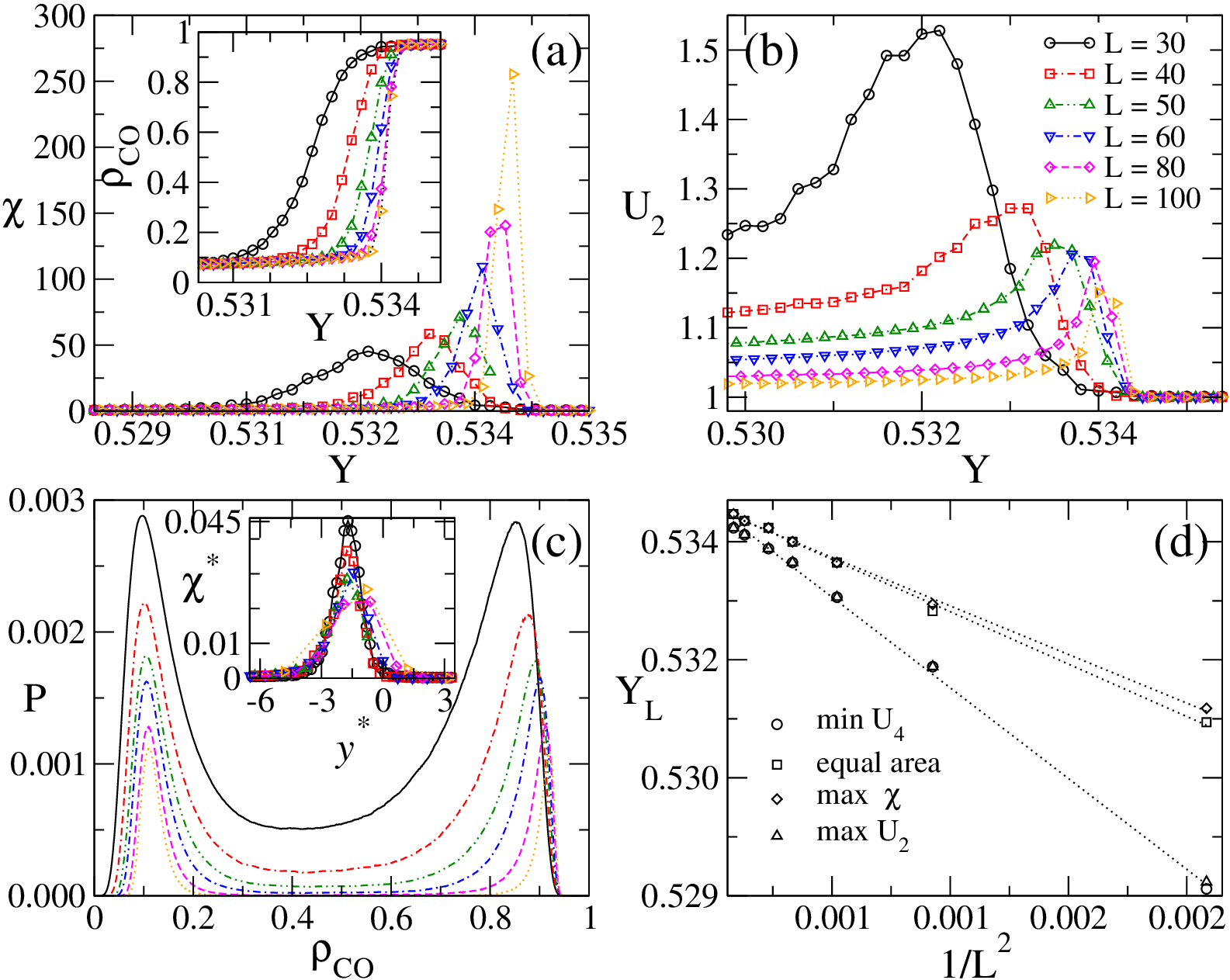}
\caption{\footnotesize{(Color online)  
The ZGB model with desorption for distinct system sizes $L$
and $k=0.02$.
(a) The order parameter variance $\chi$ (and order-parameter 
$\rho_{CO}$, inset) versus the creation rate $Y$. 
(b) The cumulant $U_2$ versus $Y$ and (c) the equal-area probability 
distribution of $\rho_{CO}$ (inset shows the data collapse
by writing ${\chi}^*=\chi/L^2$ and $y^*=(Y-Y_0)L^2$). 
(d) The scaling plots of $Y_L$'s as function of $1/V = L^{-2}$. }}
\label{zgbfss1b}
\end{figure}

We run the simulations for $t=10^7$ MC steps and
appropriate averages are evaluated after discarding the initial
$10^6$ MC steps. 
The results for $k=0.02$ are summarized in Fig. \ref{zgbfss1b}. 
We clearly see typical trends of a discontinuous phase transition:
the variation of $\rho_{\mbox{\scriptsize CO}}$ with $Y$ (inset of Fig.
\ref{zgbfss1b} (a)) becomes steeper as the system size ($L$) increases,
taking place in narrower intervals $\Delta Y$, which tend to agree
with the maximum regions of $\chi$ (Fig. \ref{zgbfss1b} (a)) and
$U_2$ (Fig. \ref{zgbfss1b} (b)) for increasing $L$.
Also, the overlap between the full $P_V(\rho_{\mbox{\scriptsize CO}})$
at the equal area condition (Fig. \ref{zgbfss1b} (c)) (and as it 
should be expected, the minimum of $U_4$, but not shown here)
decreases with $L$. 
The position of the peaks and minima are found to scale as
$1/V = L^{-2}$, from which we obtain the estimate (Fig. \ref{zgbfss1b}
(d)) $Y_0=0.5342(1)$
(max. $\chi$),  $Y_0=0.5343(1)$ (equal area $P_V^{(C)}$'s) and
$Y_0=0.5343(1)$ (min. $U_4$ and max. $U_2$).
Note the great concordance of the different measures for $Y_0$.
Finally, the inset in Fig. \ref{zgbfss1b} (c) shows a data collapse
using ${\chi}^* = \chi/L^2$ and $y^* = (Y-Y_0) L^2$, confirming the
correct scaling with $V = L^2$.

{\em Majority vote model with inertia}:
The majority vote (MV) is one of the simplest nonequilibrium up-down
symmetric $Z_2$ model exhibiting an order-disorder phase transition
\cite{mario92}. 
To each site $i$ of a network --- whose vicinity is formed by its
$k$ nearest neighbors --- corresponds a spin variable $\sigma_i = \pm 1$. 
In the original model, with probability $1-f$ the value of $\sigma_i$
turns to that of the majority of the sites in the $i$ vicinity. 
By increasing the misalignment parameter in the interval
$0 \leq f \leq 1/2$, the system undergoes a continuous nonequilibrium
phase transition from an ordered to a disordered state
\cite{mario92,chen1,pereira}. 

It has been proved \cite{chen1} that the continuous phase transition
becomes first-order when a term depending on the local density is
included in the MV dynamics, an inertial effect (for a similar
mechanism in a cellular automata model, see \cite{klaus}).
Subsequent works have shown that even a partial inertia (i.e., for only
a fraction of the sites) can change the transition characteristic
depending on the inertia strength $\theta$ \cite{pedro}, clarifying
the necessary ingredients to originate the DNPT \cite{jesus}. 
For low $\theta$ and $k$, the phase transition is continuous, whereas
it is discontinuous for both parameters high enough.

Very recently it has been verified \cite{jesus} that the same sort
of continuous to discontinuous change --- observed in networks ---
can take place in regular square lattices.
At each time step, a randomly chosen $\sigma_i$ is flipped
($\sigma_i \rightarrow -\sigma_i$) according to the probability (adapted
from \cite{chen1} for such regular topology)
\begin{equation}
w(\sigma_i) =
\frac{1}{2} + \left(f - \frac{1}{2}\right) \sigma_{i} \,
\mbox{sign} \! \! \! \,
\left[\frac{(1-\theta)}{k} \sum_{j=1}^{k} \sigma_j + \theta \, \sigma_i
\right],
\label{eq8}
\end{equation} 
where for a discontinuous phase transition a minimal of $k=20$ is required.

\begin{figure}[t]
\centering
\includegraphics[scale=0.31]{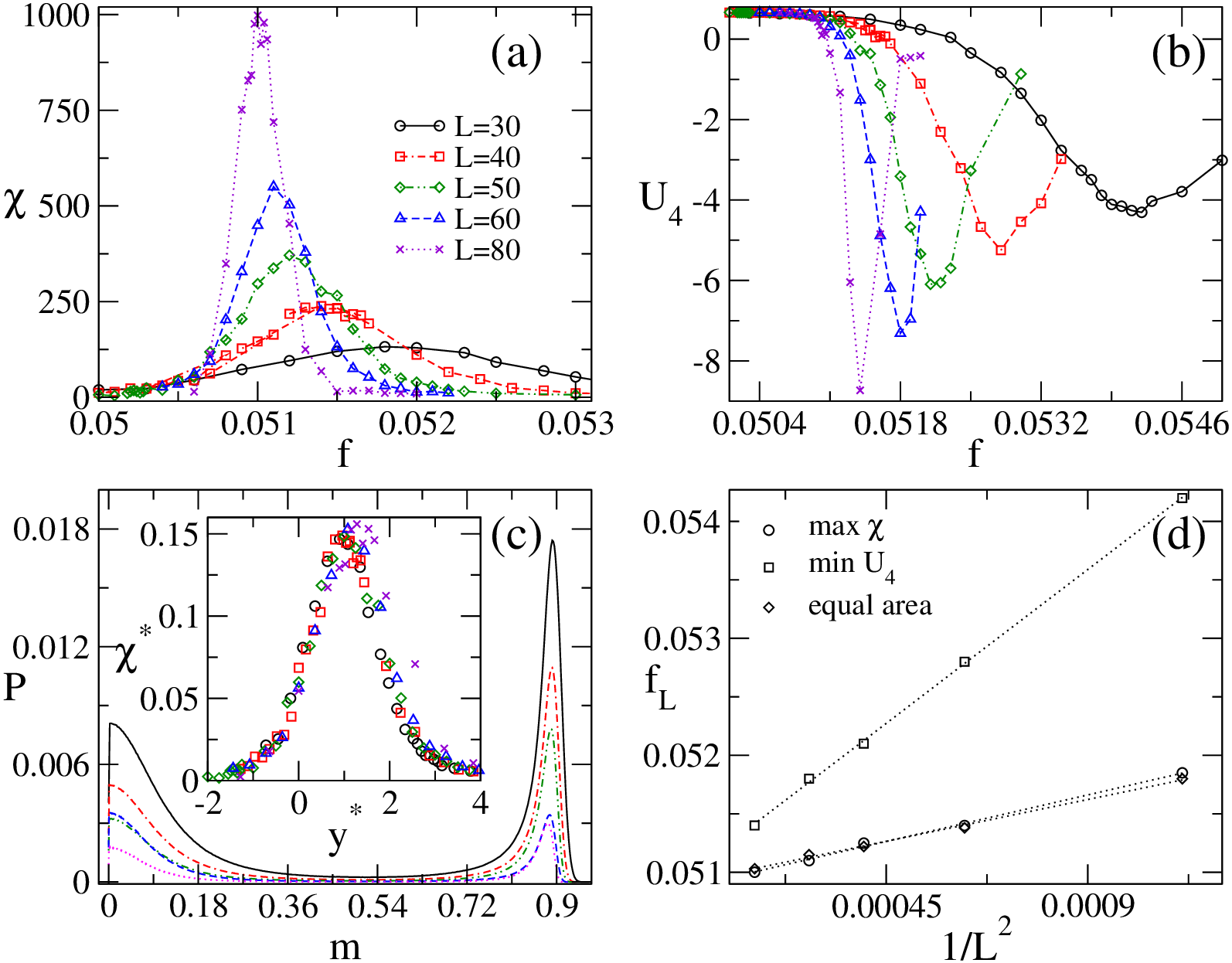}
\caption{(Color online) The majority vote model with inertia 
for distinct system sizes $L$, $\theta=0.375$ and $k=20$.
(a) The order-parameter variance $\chi$ and (b) $U_4$ 
versus $f$.
(c) The equal area probability distribution of $m$ 
(inset, the data collapse by writing 
$\chi^*=\chi/L^{2}$ and $y=(f-f_0)L^{2}$).
(d) The scaling plots of $f_L$'s versus $1/V = L^{-2}$.}
\label{fig2}
\end{figure}


In Fig. \ref{fig2} we show the simulation results for $\theta=0.375$ and
$k=20$.
Here also all the quantities scale with the inverse of the system 
volume, as observed for the maximum of $\chi$ (Fig. \ref{fig2} (a)),
minimum of $U_4$ (Fig. \ref{fig2} (b)) and equal area distribution of $m$
(Fig. \ref{fig2} (c)).
Such quantities provide the estimates for $f_0$ (Fig. \ref{fig2} (d)),
$0.0509(1)$, $0.0510(1)$, and $0.0509(1)$, respectively.
The data moreover collapse by setting $\chi^* = \chi/L^{2}$ and
$y = (f-f_0) L^{2}$ (inset of Fig. \ref{fig2} (c)).


{\em Competitive contact process in bipartite sublattices}:        
Lastly, we consider a system with competitive interactions in bipartite 
sublattices, introduced originally in \cite{martins}.
It exhibits two types of transitions:
(i) spontaneous symmetry-broken phase transitions, between two
active states, and
(ii) a continuous absorbing phase transition, whose critical behavior and
universality class, DP, are not affected by the particle diffusion
\cite{oliveira17}.
Nonetheless, by requiring a minimal occupied neighborhood to create
a new particle (a restrictive interaction) it gives rise to distinct
sorts of DNPT \cite{salete}.

The model dynamics is described as the following. 
A particle in a given sublattice $j$ (A or B) creates autocatalitically a
new particle with rate $\lambda_1 \, n_{1 j}/4$ ($\lambda_2 \, n_{2 j}/4$)
in one of its nearest $s=1$ (next-nearest $s=2$) neighbor empty sites.
Here, $\lambda_s$ is the creating strength parameter and $n_{s j}$
denotes the number of particles in the corresponding $s$ neighborhood of
the considered site in sublattice $j$.
A transition requires $n_{s j} \ge s$ adjacent particles
(the condition $n_{1 j} \ge 1$ is similar to that of the original contact
process model).
This slightly modification \cite{martins} leads to the
appearance of three coexistence lines, instead of critical ones. 
Additionally, an inhibition term depending on the local density, in the
form $\mu \, n_{1 j}^2$, is included.
This favors unequal sublattice populations \cite{martins}.
If $\mu=0$, one recovers the traditional case in which a particle is
spontaneously annihilated with rate $1$. 
Besides the typical absorbing (ab) and active-symmetric (as),
we observe a stable active-asymmetric (aa) phase for intermediate 
control parameters values.
The  ab--as and  aa--as transition lines are discontinuous (continuous)
for $\lambda_1$ low (large).
The finite size scaling for the absorbing first-order transition has
been studied in \cite{fss15}.

\begin{figure}[!t]
\includegraphics[scale=0.31]{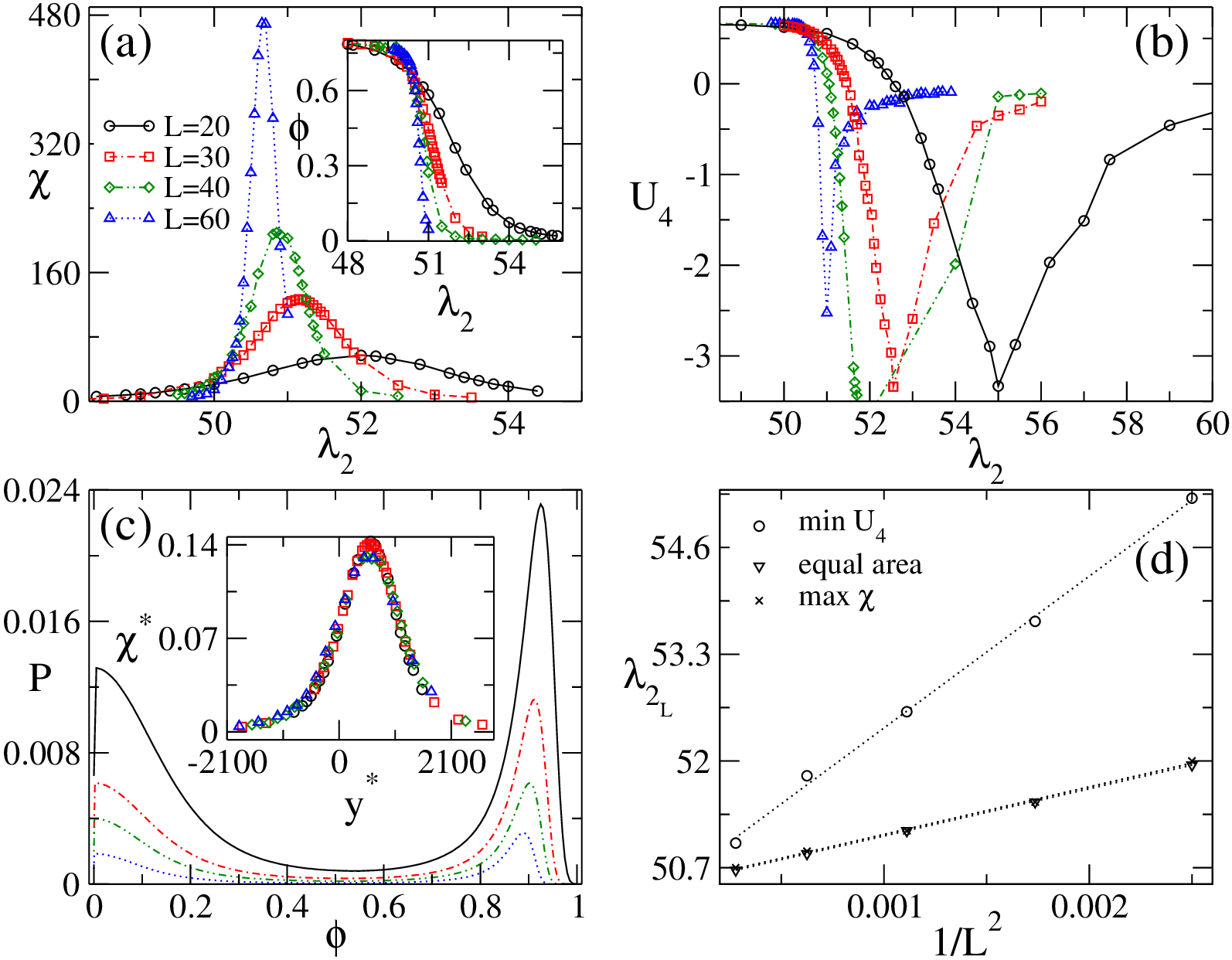}
\caption{\footnotesize{(Color online)
The competitive contact process in bipartite sublattices for the
$aa-as$ transition, distinct $L$'s, $\lambda_1 = 0.5$
and $\mu=1$.
(a) The order-parameter variance $\chi$ and the order-parameter 
$\phi$ (inset) versus $\lambda_2$. 
(b) The reduced cumulant $U_4$ versus $\lambda_2$.
(c) The equal-area probability distribution of $\phi$
(inset, data collapse writing $\chi^*=\chi/L^2$ and
$y^*=(\lambda_2-\lambda_{20})L^2$). 
(d) The scaling of distinct pseudo-transition points $(\lambda_{2})_L$ 
versus $1/V = L^{-2}$.}}
\label{fig3}
\end{figure}

On the other hand, the aa--as  line is not of absorbing type,
furthermore displaying spontaneous breaking of symmetry.
To locate it, we can evaluate the sublattice particle densities
$\rho_j$.
In the as phase $\rho_A = \rho_B \neq 0$, whereas in the aa phase,
$\rho_A \neq \rho_B$ and $\rho = \rho_A + \rho_B \neq 0$.
Since sublattices are unequally populated in the aa phase, a natural
choice for an order parameter is the difference of sublattice densities,
or $\phi = |\rho_A-\rho_B|$.
Unlike the as phase, in which $\phi=0$, in the aa phase $\phi \neq 0$.

In Fig. \ref{fig3} (a)--(c) we show variance $\chi$, $U_4$, and equal
area distribution of $\phi$, for the aa-as
phase transition with $\lambda_1=0.5$, $\mu = 1$ and different $L$'s.
Hence, $\lambda_2$ is the control parameter.
Previous calculations \cite{salete} show that the symmetry-breaking
transition occurs in the vicinity of $\lambda_2=50.96$ (evaluated
for $L=80$).
Here, the finite size analysis leads to much more precise estimations.
As the previous examples, the pseudo-transition points
$(\lambda_{2})_L$'s, Fig. \ref{fig3} (d), scale with $1/V = L^{-2}$,
from which we obtain ${(\lambda_{2})}_0=50.55(2)$ (minimum $U_4$),
$50.51(1)$ (equal area) and $50.53(2)$ (maximum $\chi$).
Once more, the collapse (inset of Fig. \ref{fig3} (c)) is very good.


Summing up, we have developed a phenomenological FSS for distinct 
types of DNPT, supposing only they correspond to stable steady states,
for which the global balance, Eq. (\ref{eq:global-balance}), holds true. 
As examples, we have considered three distinct processes: 
(1) transitions between low-density (high active) and high-density
(low active) phases, in the context of the ZGB model with desorption;
(2) order-disorder phase transitions induced by inertial effects, 
discussing the majority vote model with inertia; and 
(3) symmetry-breaking transitions between active phases, for a competitive 
contact process in bipartite sublattices. 
In all these cases, the finite-size behaviors are fully described by the 
present phenomenological theory.

We believe the results here derived will help to put the study and
understanding of generic (nonequilibrium) discontinuous phase transition
on more firm basis.

\acknowledgments
We acknowledge CNPq, CT-Infra, and Fapesp for research grants.

\bibliographystyle{apsrev}

\newpage

\end{document}